\documentclass[reprint,amsmath,amssymb,aps,prl]{revtex4-2}
\usepackage[latin9]{inputenc}
\usepackage{amsmath}
\usepackage{graphicx}

\makeatletter
\usepackage{xcolor}
\usepackage{lipsum}
\graphicspath{{./Figures/}}

\usepackage{comment}
\usepackage{float}

\makeatother

\begin{document}
\title{A Memory-Based Approach to Model Glorious Uncertainties of Love}
\author{Aarsh Chotalia}
\thanks{These authors contributed equally to this work}
\author{Shiva Dixit}
\thanks{These authors contributed equally to this work}
\author{P. Parmananda}
\affiliation{Department of Physics, Indian Institute of Technology Bombay, Powai,
Mumbai $400$ $076$, India.}
\begin{abstract}
We propose a minimal yet intriguing model for a relationship between
two individuals. The feeling of an individual is modeled by a complex
variable and hence has two degrees of freedom \citep{jafari2016layla}.
The effect of memory of other individual's behavior in the past has
now been incorporated via a conjugate coupling between each other's
feelings. A region of parameter space exhibits multi-stable solutions
wherein trajectories with different initial conditions end up in different
aperiodic attractors. This aligns with the natural observation that
most relationships are aperiodic and unique not only to themselves
but, more importantly, to the initial conditions too. Thus, the inclusion
of memory makes the task of predicting the trajectory of a relationship
hopelessly impossible. 
\end{abstract}
\date{\today}

\maketitle
Human interactions can take place over a vast range of timescales.
Of those, romantic relationships are the most intimate i.e. strongly
coupled and generally long-standing. One approach to model such relations
is to consider a dynamical system consisting of a large number of
psychosomatic variables such as hormonal levels and employ neuropsychology-informed
observations \citep{rubin1970measurement} to model the coupling and
the effect of environment. Our approach in this work, however, is
to consider a minimal model which is able to generate complex dynamics
akin to what is observed in the real world.

Depending upon the nature of the coupling, a dynamical system may
show a wide range of behaviors such as amplitude death, oscillation
death, limit cycles, and chaos as the system parameters vary~\citep{HAN201673,ponrasu2018conjugate,taher2018multistability,crowley1989experimental,shajan2021enhanced}.
While modelling love dynamics has a long history starting from linear
models \citep{strogatz2018nonlinear,rinaldi1998love}, there have
been several recent attempts to model the dynamics of a romantic relationship
with the inclusion of non-linearities \citep{jafari2016layla,owolabi2019mathematical,ozalp2012fractional,DENG201713,kumar2021complex}.
For a dynamicist, the prime challenge is to quantify love as well
as the coupling interaction between the lovers. Naturally, a sensible
model must incorporate memory. The role of memory on affect has been
discussed by psychologists\citep{blaney1986affect}. The models in
\citep{ozalp2012fractional,kumar2021complex} and \citep{bielczyk2012delay,DENG201713}
have included memory effects using fractional ordered derivatives
and time-delayed equations respectively. In this work, the effect
of memory is encapsulated by a conjugate coupling~\citep{yadav2019universal,Singla2011ExploringTD,karnatak2009synchronization,ponrasu2020aging,karnatak2014conjugate}
between the two lovers' feelings.

Typically, the variables in a dynamical systems are taken to be real
numbers as they generally correspond to some real-world observable.
A straightforward way to quantify feeling is to assign each of the
two lovers a real number $l_{i}~(i=1,2)$. The sign of $l_{i}$ would
determine whether the feeling is positive (i.e. love) or negative
(i.e. hatred), while $\left|l_{i}\right|$ would represent the intensity
of the feeling. However, the types of behavior of dynamics are very
limited for a two-dimensional phase space. Since we have two individuals,
we must assign two real numbers to each lover. We, therefore, promote
$l_{i}$ to a complex number $L_{i}$ $\in\mathbb{C}\cong\mathbb{R}^{2}$
(henceforth referred to as ``complex love") of which real and imaginary
parts are two real degrees of freedom making the phase space effectively
four-dimensional. This, as we shall see in the following paragraphs,
makes the task of incorporating conjugate coupling easier and more
natural.

\begin{figure}
\vspace{10pt}
 \includegraphics[width=0.45\textwidth]{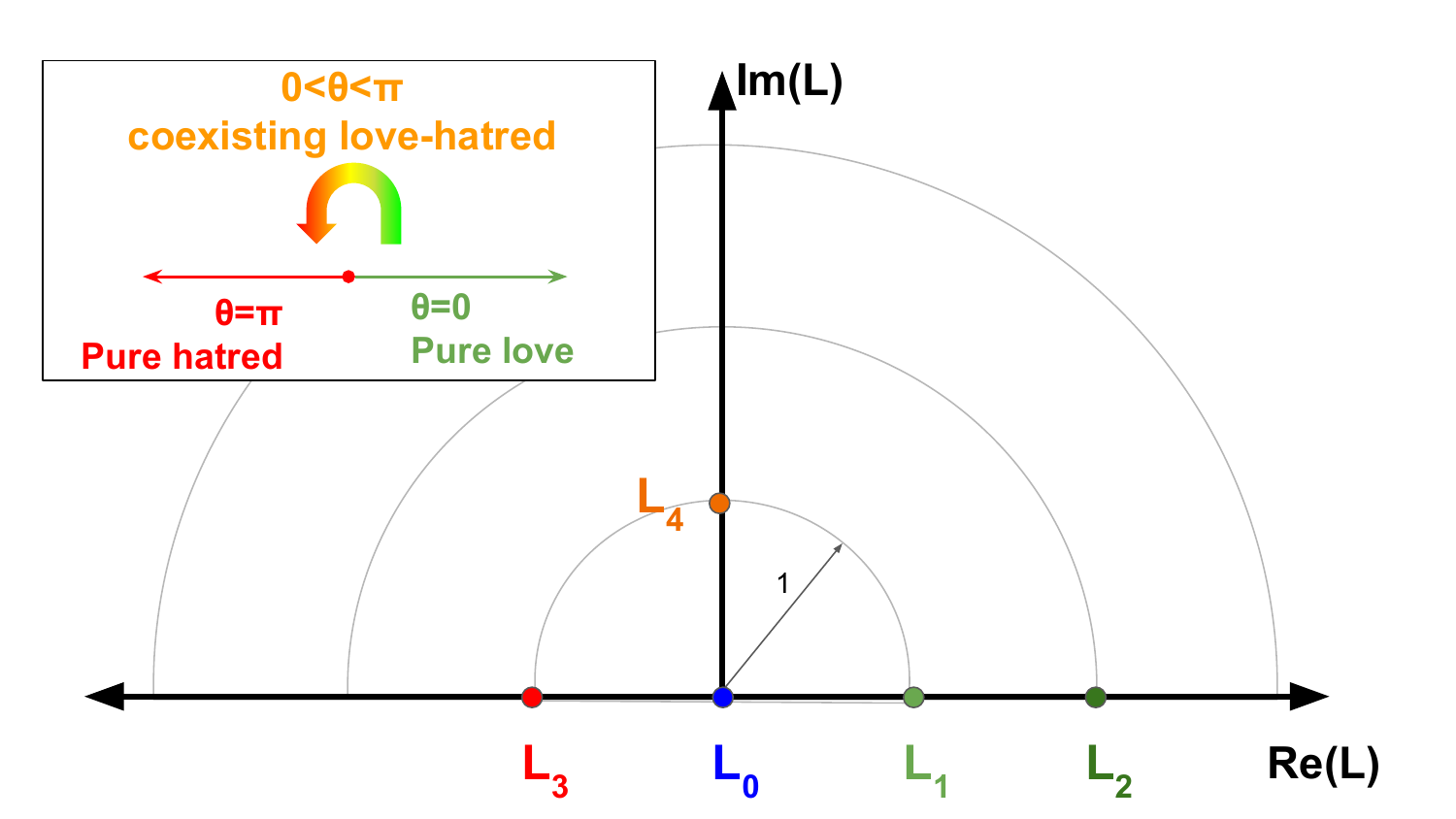} \caption{(Color online) Love on the Argand plane: Interpretation of a complex
love. The points $L_{0}$ to $L_{4}$ are examples of complex love.
$L_{0}=0+0i$: apathy, $L_{1}=1+0i$: one unit pure love, $L_{2}=2+0i$:
two units pure love, $L_{3}=-1+0i$: one unit pure hatred and $L_{4}=0+i$:
coexisting half unit love and half unit hatred.}
\label{figure1} 
\end{figure}

The complex love number is to be interpreted in the following manner,
which was first given by \citep{jafari2016layla}. Love and hatred
can coexist and the phase of the complex love determines the fraction
of love. A phase of zero corresponds to a state of pure love while
a phase of $\pi$ corresponds to a state of pure hatred. Any intermediate
phase $\theta\in(0,\pi)$ would correspond to a state with co-existing
love-hatred with a fraction of love $f_{l}=\theta/\pi$ e.g. a phase
of $\pi/2$ would mean the person has an equal amount of love as hatred
towards the other. Figure \ref{figure1} shows some examples of possible
complex loves. It should be noted that, under this interpretation,
a complex love with a phase of $\pi+\theta$ is to be considered equivalent
to one with a phase of $\pi-\theta$.

While the time delay introduces coupling between the same variable
at different times, conjugate coupling couples different variables
at the same time. The latter serves as an indirect method to introduce
a time delay without making the equations explicitly non-local in
time. This accords with Takens' embedding theorem \citep{takens1981dynamical}
which states that a chaotic attractor may be reconstructed by using
a time-delayed sequence of one component. The conjugate coupling has
been used to capture time-delay by \citep{karnatak2014conjugate,Manju,dasgupta2010suppression}
in other settings. Mathematically, a dynamical system $\{\mathbf{x}_{1},\mathbf{x}_{2},...,\mathbf{x}_{n}\}$
where each $\mathbf{x}_{i}\in\mathbb{R}^{m}$ having a conjugate coupling
evolves as

\begin{equation}
(\dot{\mathbf{x}_{i}})_{j}=\left(\mathbf{f}_{i}(\mathbf{x}_{i})\right)_{j}+F_{ij}((\mathbf{x}_{i})_{j},(\mathbf{x}_{k})_{l})
\end{equation}

where $i,k=1,..,n;j,l=1,...,m;k\neq i;l\neq j$.

In other words, the coupling term of the evolution of a given component
of a given vector depends on itself and the complementary components
of the other vectors. In particular, for a system of two 2-d vectors
let $j^{th}$ component of $\mathbf{x}_{i}$ be $x_{ij}$ where $i,j=1,2$,
then $x_{11}$ is coupled to $x_{22}$ whereas $x_{12}$ is coupled
to $x_{21}$ and vice-versa.

We name the two lovers ``Layla" and ``Majnun" after the central
characters of the classic Persian love story ``Layla-e-Majnun".
Let $L(t)$ ($M(t)$) be the complex love Layla (Majnun) has for Majnun
(Layla). The model given by Jafari et al. \citep{jafari2016layla}
for the temporal evolution of the relationship between Layla and Majnun
is: 
\begin{align}
\frac{dL}{dt} & =a+M^{2}+c_{1}L\label{j_equation1}\\
\frac{dM}{dt} & =b+L^{2}+c_{2}M\label{j_equation2}
\end{align}
wherein the authors identify the parameters $a$ and $b$ as ``environmental
factors" for Majnun and Layla respectively. The environmental factor
indicates the tendency of an individual to engage in a relationship
due to their sociocultural environment. In the case of the Layla-Majnun
story, in most versions, Layla is married to someone else, and therefore
$a$ is negative while Majnun is a young vagabond poet, looking for
love, and therefore $b$ is positive.

Our proposed model differs from the previous one in the final terms:

\begin{align}
\frac{dL}{dt} & =a+M^{2}+c_{1}(L-iM)\label{model_equation1}\\
\frac{dM}{dt} & =b+L^{2}+c_{2}(M-iL)\label{model_equation2}
\end{align}

The final terms indicate the conjugate coupling between $M$ and $L$.
Note that when equations (\ref{model_equation1},\ref{model_equation2})
are split into real and imaginary parts, the real part of $L$($M$)
gets coupled to the imaginary part of $M$($L$) and vice-versa. As
described earlier, this incorporates the memory of each other's feelings
in the past into the evolution of the relationship. For this reason,
we term $c_{1}$ and $c_{2}$ as cross-memory parameters. One may
easily deduce that the system is Hamiltonian if and only if $c_{1}+c_{2}=0$
with 
\begin{equation}
H=bL-aM+\frac{ic}{2}\left(L^{2}-M^{2}\right)-cML+\frac{L^{3}-M^{3}}{3}
\end{equation}
where $c=c_{1}=-c_{2}$.

The fixed points of the system satisfy the following quartic equations:

\begin{align}
 & \begin{aligned}L^{4}-2ic_{2}L^{3}+ & \left(2b-c_{2}^{2}+ic_{1}c_{2}\right)L^{2}+\left(-2ibc_{2}+2c_{1}c_{2}^{2}\right)L\\
 & \qquad+ac_{2}^{2}+b^{2}+ic_{1}c_{2}b=0
\end{aligned}
\\
 & \begin{aligned}M^{4}-2ic_{1}M^{3}+ & \left(2a-c_{1}^{2}+ic_{1}c_{2}\right)M^{2}+\left(-2iac_{1}+2c_{1}^{2}c_{2}\right)M\\
 & \qquad+bc_{1}^{2}+a^{2}+ic_{1}c_{2}a=0
\end{aligned}
\end{align}

The Jacobian of (\ref{model_equation1},\ref{model_equation2}) in
the phase space $(L_{r},L_{i},M_{r},M_{i})$, where subscripts $r$
and $i$ denote the real and imaginary parts respectively, is given
by:\\

\begin{equation}
J=\begin{bmatrix}c_{1} & 0 & 2M_{r} & -2M_{i}+c_{1}\\
0 & c_{1} & 2M_{i}-c_{1} & 2M_{r}\\
2L_{r} & -2L_{i}+c_{2} & c_{2} & 0\\
2L_{i}-c_{2} & 2L_{r} & 0 & c_{2}
\end{bmatrix}\label{jack}
\end{equation}

We simulated the equations (\ref{model_equation1},\ref{model_equation2})
for different initial conditions as well as for different environmental
factors and cross-memories using \textit{MATLAB ver. $R2021b$} with
the default differential equation solver ode$45$, which uses an optimum-adaptive
step size. The initial conditions $(L_{r}(0),L_{i}(0),M_{r}(0),M_{i}(0))$
were drawn uniformly from the interval $[-1,1]$.

\begin{figure}[h!]
\vspace{10pt}
 \includegraphics[width=0.45\textwidth]{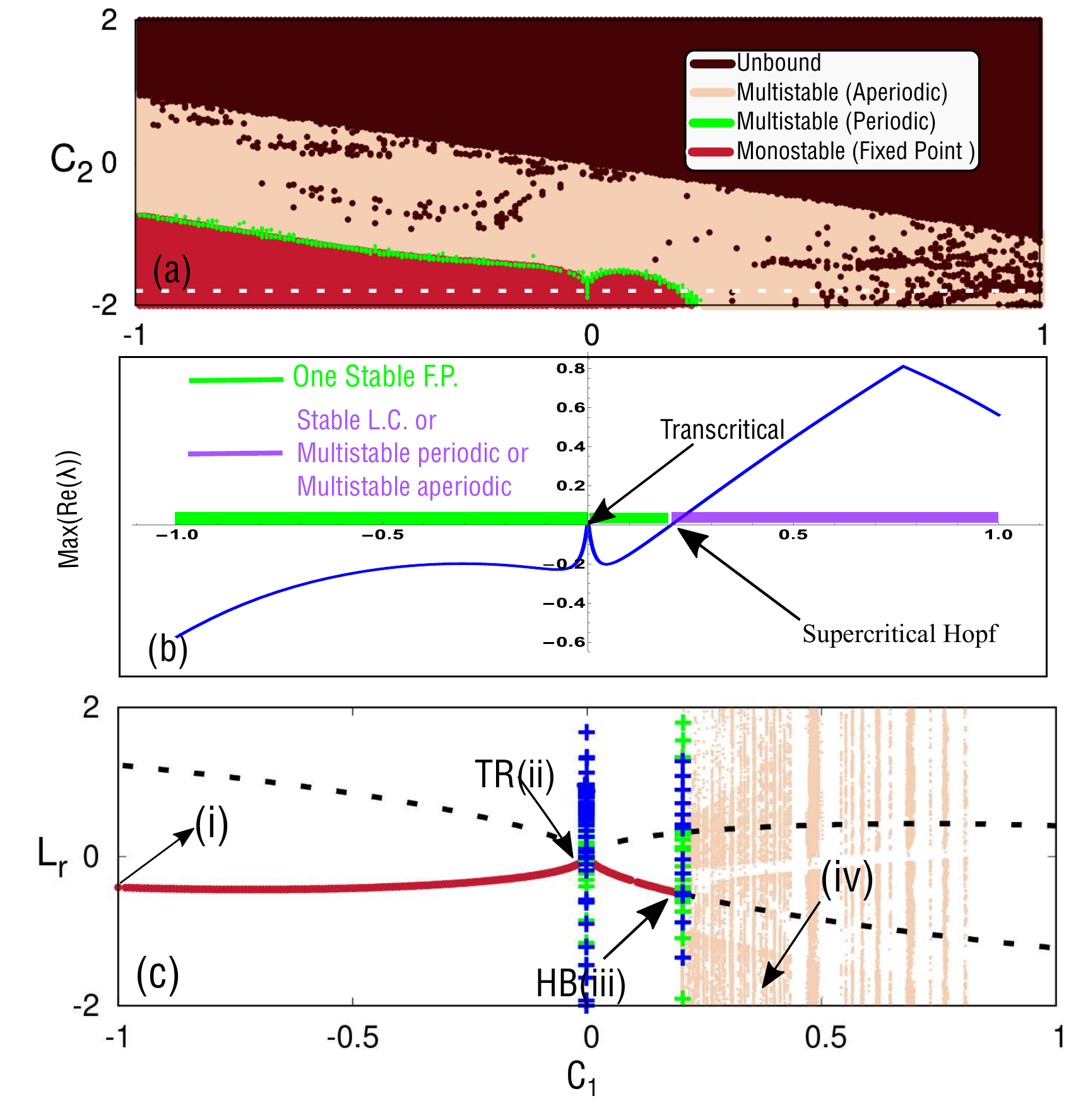} \caption{(Color online)(a) The stability of the system in $(c_{1},c_{2})$
plane: The environmental factors $a$ and $b$ are fixed to be $-1$
and $1$ respectively. The cream region is where multiple aperiodic
orbits exist for different initial conditions. (b) Largest real part
of eigenvalue of the Jacobian and (c) The bifurcation diagram produced
using the software XPPAUT~\citep{ermentrout2007xppaut} as $c_{1}$
is varied between $-1$ and $1$ at fixed $c_{2}=-1.8$ (i.e., along
the white dashed horizontal line in (a)). For details see the text.}
\label{figure2} 
\end{figure}

\begin{figure}
\vspace{10pt}
 \includegraphics[width=0.45\textwidth]{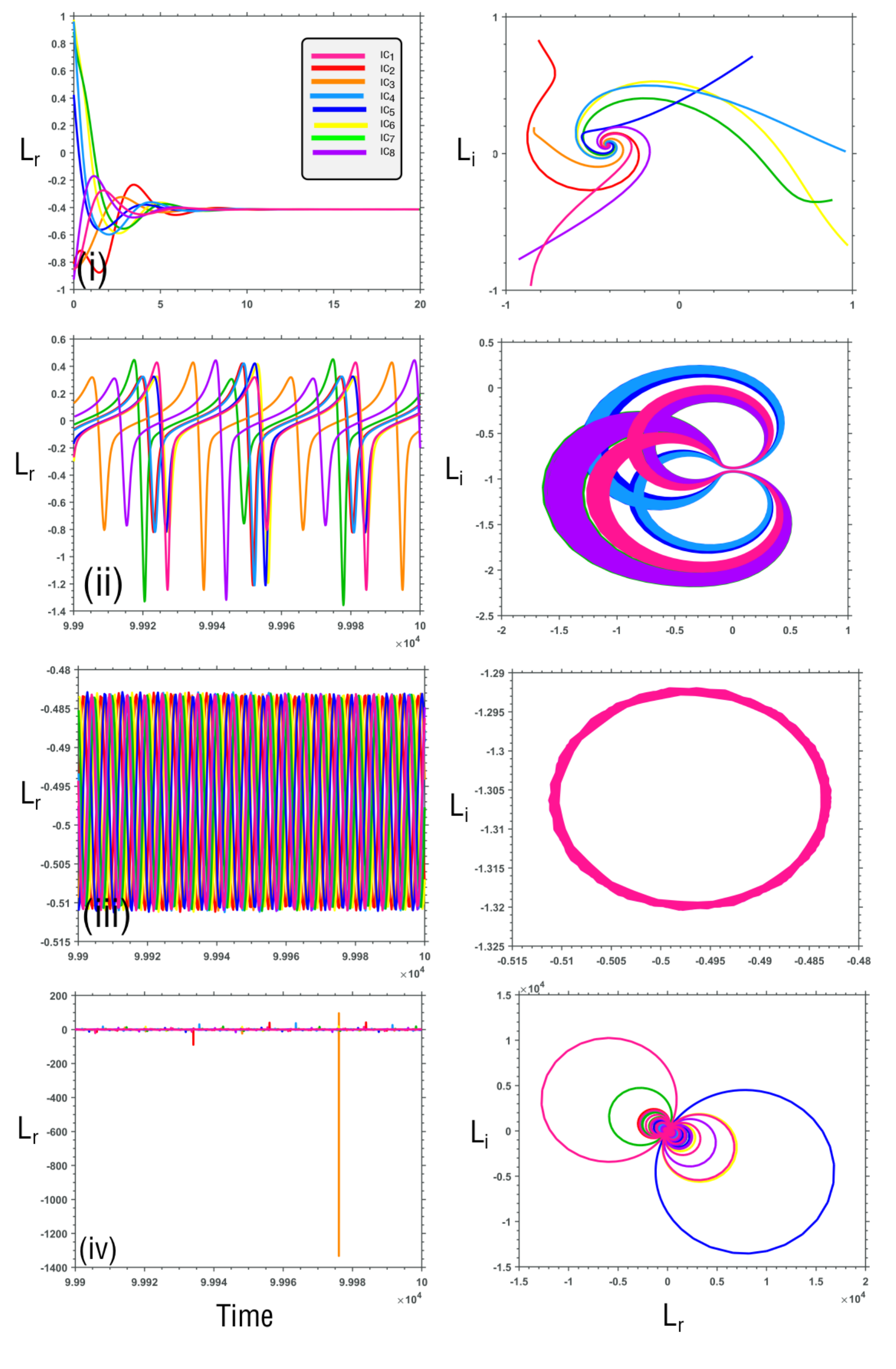} \caption{(Color online) Time-series corresponding to Figure \ref{figure2}
(b) of $L_{r}$ and the trajectories in the Layla plane (i.e., $L_{r}$-$L_{i}$
plane) for four different values of $c_{1}=(i)-1.0,(ii)0.0,(iii)0.2\text{ and},(iv)0.4$
with fixed $(a,b,c_{2})=(-1,+1,-1.8)$. Each subfigure (i)-(iv) has
time-series and trajectories corresponding to eightifferent initial
conditions having each of the four variables drawn uniformly from
$[-1,1]$.}
\label{figure3} 
\end{figure}

Firstly, we consider the case with fixed environmental factors ($a=-1$
and $b=1$) and varying cross-memories $c_{1}$ and $c_{2}$ which
are crucial model parameters that define the relationship. Our key
interest is to identify the emergent collective dynamics due to the
interplay of the cross memory strengths. Figure \ref{figure2} (a)
reveals a host of behaviours as the $(c_{1},c_{2})$ space is traversed
through $[-1,1]\times[-2,2]$. Noticeably, the $c_{1}=-c_{2}$ (white
solid line) line on which the system is Hamiltonian also demarcates
the regions of bound and unbound trajectories. The stability of the
system can be understood from the eigenvalues of the Jacobian (equation
\ref{jack}) at the fixed points. Figure \ref{figure2} (b) shows
the largest (of four) real parts of the eigenvalues as $c_{1}$ is
varied from -1 to 1 keeping $c_{2}=-1.8$ fixed (white dashed line
in Figure \ref{figure2}(i)). The two green intervals $[-1,0)\cup(0,0.2)$
represent the values of $c_{1}$ for which the system has a stable
fixed point as $Re(\lambda_{max})$ is negative in these intervals.
At $c_{1}=0$, two solutions exchange their stabilities due to a transcritical
bifurcation. When $c_{1}$ approaches a value of about 0.2, the largest
eigenvalue becomes positive and the stable solution, as shown in Figure
\ref{figure3} (iii), gives rise to stable limit cycle encircling
a stable fixed point. This indicates to emergence of a supercritical
Hopf bifurcation. Figure \ref{figure2} (c) is the bifurcation diagram
for $L_{r}$ for the same range of $c_{1}$: The fixed stable point
(red) bifurcates into multiple stable limit cycles (green and blue)
via a transcritical bifurcation at $c_{1}=0$ before again becoming
stable (red). At $c_{1}\approx0.2$, stable fixed point (red), through
a supercritical Hopf bifurcation, gives rise to stable limit cycle
(green) which in turn generates multiple limit cycles (blue) before
reaching a region (cream) where multiple aperiodic trajectories corresponding
to different initial conditions reside. As shown in Figure \ref{figure2}(a),
once $c_{1}$ exceeds $-c_{2}$, the basins for chaotic attractors
vanish and all trajectories become unbounded (brown). This emphasises
the need of having cross-memory since at $(c_{1},c_{2})=(0,0)$, the
system lives at the edge between multistable aperiodic attractors
and the unbounded region. Similar bifurcations for varied $c_{2}$
and fixed $c_{1}$ have been presented in the supplementary material.

To illustrate the richness of the dynamics as cross-memory is varied,
Figure \ref{figure3} shows the time-series of $L_{r}$ and the trajectories
in the Layla plane (i.e., $L_{r}$-$L_{i}$ plane) for four different
values of $c_{1}$ in the bifurcation curve Figure \ref{figure2}
(c) i.e., with fixed $(a,b,c_{2})=(-1,+1,-1.8)$ . Each subfigure
(i)-(iv) shows eight time-series and trajectories corresponding to
eight different initial conditions having each of the four variables
drawn uniformly at random from $[-1,1]$. (i) At $c_{1}=-1$, the
system has a stable fixed point $(L_{r}^{*},L_{i}^{*},M_{r}^{*},M_{i}^{*})=(-0.41,0.07,0.58,-0.44)$.
(ii) At $c_{1}=0$ and (iii)$c_{1}=0.2$, the system has multiple
limit cycles and a single limit cycle, respectively. (iv) At $c_{1}=0.4$
multiple aperiodic trajectories exist corresponding to different initial
conditions. The periodic solutions reflect a typical stable relationship
wherein the partners' emotions change slowly and predictably. On the
contrary, the sporadic spikes in the aperiodic solutions are indicative
of abrupt emotional fluctuations one experiences sometimes. A real
life couple may indeed have time-dependent cross-memories that can
lead to a solution that switches between periodic and aperiodic behaviour.

\begin{figure}[h!]
\vspace{10pt}
 \includegraphics[width=0.45\textwidth]{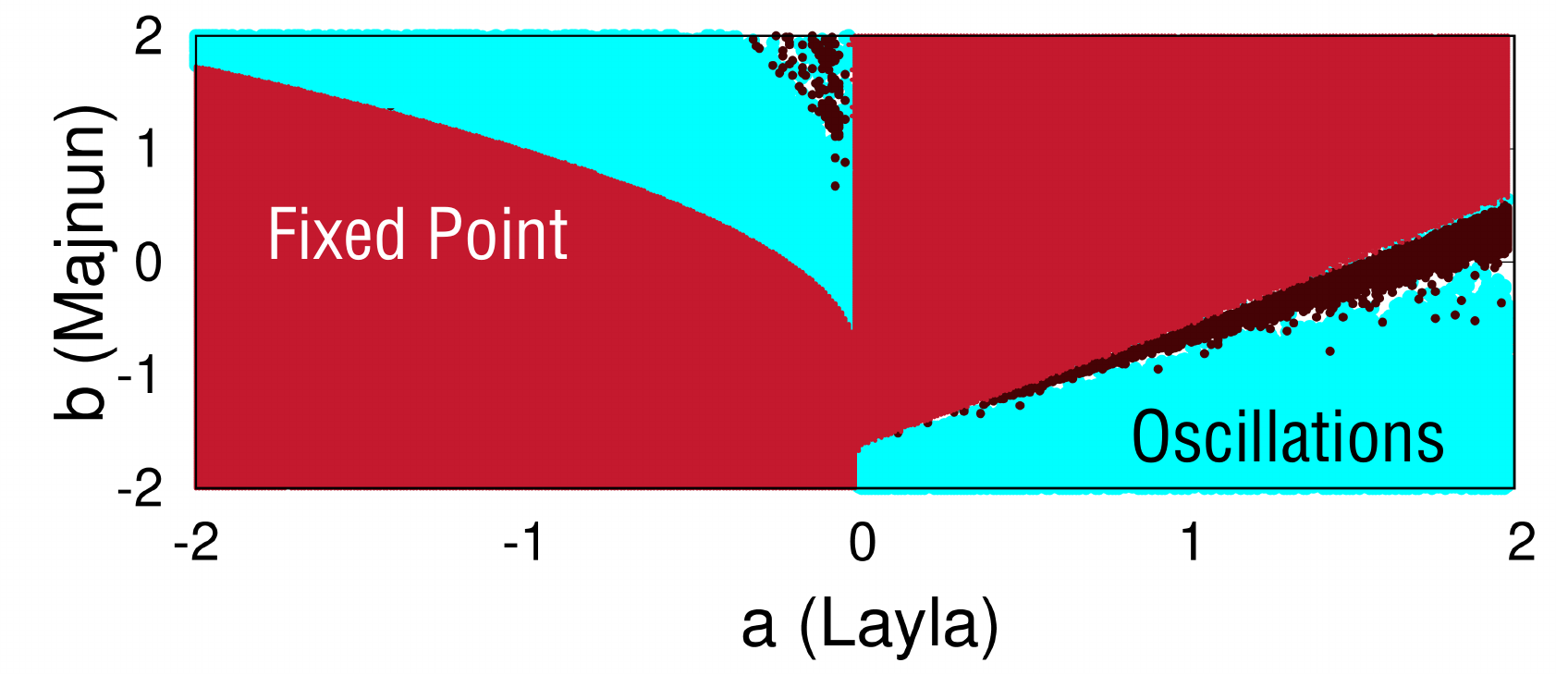} \caption{(Color online) Effect of environmental factors $a$ of Layla and $b$
of Majnun with fixed cross-memories $c_{1}=0$ and $c_{2}=-1.8$.
The red region has a stable fixed point while the cyan region exhibits
multi-stable aperiodic solutions. In the black region, the system
dynamics become unbounded for all initial conditions.}
\label{figure4} 
\end{figure}

Finally, to understand how the external environment affects the dynamics,
we varied the values of $a$ and $b$ between $-2$ and $2$. As noted
earlier, these environmental factors reflect the effect of socio-cultural
environment on an individual's tendency to love and thus serves as
an overall bias in one's dynamics (equations \ref{model_equation1},\ref{model_equation2}).
Figure \ref{figure4} shows the dynamics of the system in the parameter
space $(a,b)\in[-2,2]\times[-2,2]$ with fixed cross-memories $(c_{1},c_{2})=(0,-1.8)$.
Three types of behaviors are observed: stable fixed points (red),
multistable aperiodic solutions (cyan), and unbounded solutions (brown).

These results indicate that it is possible to obtain rich dynamics
of human relationships just by considering a minimal model. Conjugate
coupling serves to incorporate the effect of cross-memory without
explicitly introducing time-delay, which is essential in understanding
the system dynamics. Very interestingly, the parameter space contains
regions wherein there are multiple chaotic attractors corresponding
to different initial conditions. This type of multistability has been
discussed in other settings by \citep{PISARCHIK2014167,singh2022chaos,yadav2017intermittent}.
This is reminiscent of real-life scenarios: a relationship can not
only be chaotic but also unique to itself for different initial conditions.
A natural extension of this would be to consider a love triangle with
three parties involved. These kinds of studies provide the basis for
eventually understanding holistic social dynamics where the number
of individuals and types of interaction is very large.

The unenviable and seemingly impossible task of capturing the dynamics
of human emotions shows the hopelessness of prediction. It could be,
on the other hand, considered the glorious uncertainities of love.
The hope is that the interested readers would improve upon this model
in a multidisciplinary fashion in an attempt to have a basic model
that captures the human emotions at a nascent level. \bibliographystyle{apsrev4-2.bst}
\bibliography{Reference}

\end{document}


\title{Supplementary Material For:\\A Memory-Based Approach to Model Glorious Uncertainties of Love}%
\author{Aarsh Chotalia}
\thanks{These authors contributed equally to this work}
\author{Shiva Dixit}
\thanks{These authors contributed equally to this work}
\author{P. Parmananda}
\affiliation{Department of Physics, Indian Institute of Technology Bombay, Powai, Mumbai $400$ $076$, India.}

\maketitle 

\section{Bifurcation along $c_2$ axis}
\begin{figure}[h!]
    \centering
    \includegraphics[width=0.45\textwidth]{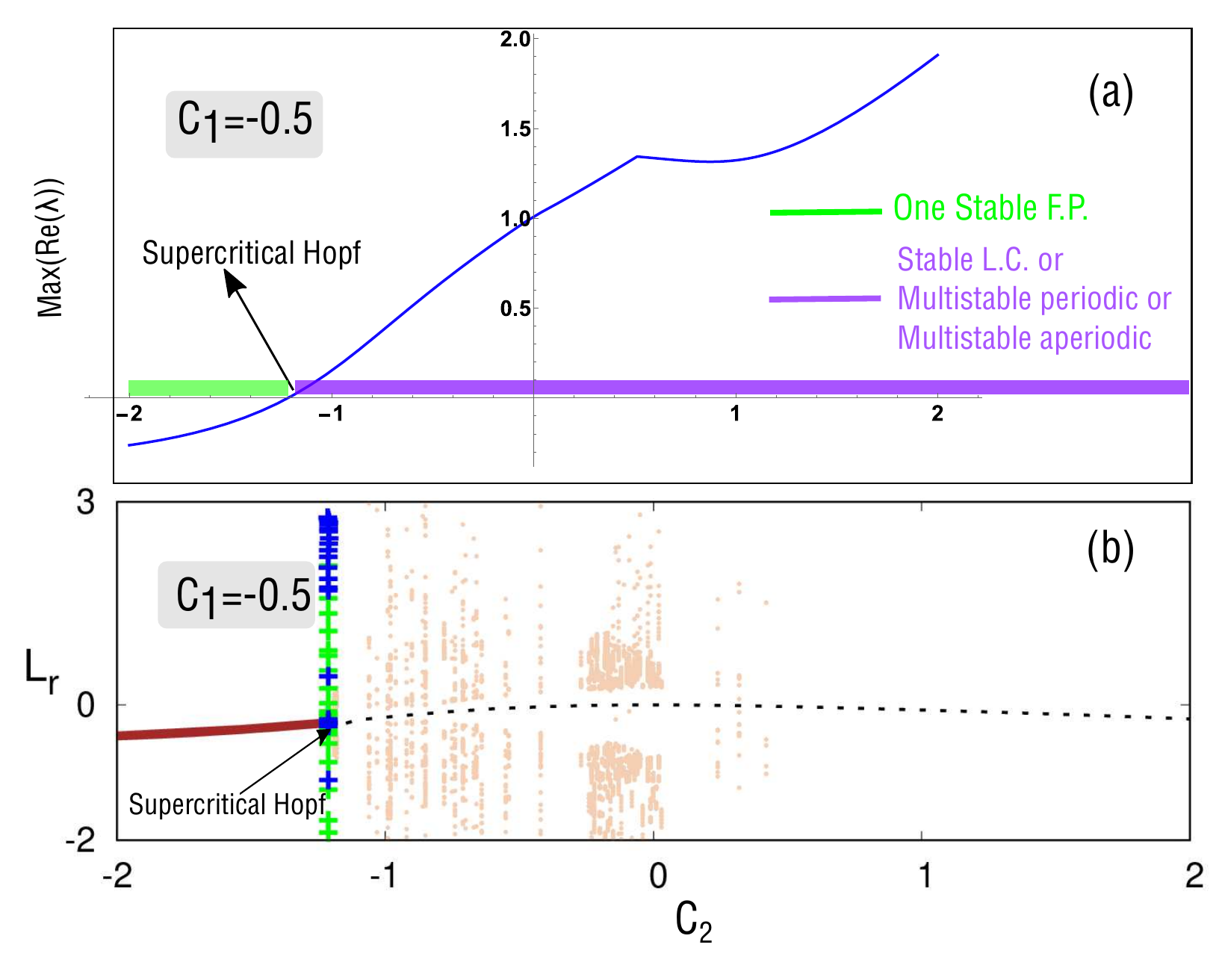}
    \caption{(a) Largest real part of eigenvalue of the Jacobian and (b) The bifurcation diagram produced using the software XPPAUT as $c_2$ is varied between $-2$ and $2$ at fixed $c_2=-0.5$}.
    \label{supfig}
\end{figure}

Figure \ref{supfig} (a) shows the largest (of four) real parts of the eigenvalues as $c_2$ is varied from -2 to 2 keeping $c_1=-0.5$ fixed. The green interval $[-2,-1.21)$ represents the values of $c_2$ for which the system has a stable fixed point as $Re(\lambda_{max})$ is negative in these intervals. When $c_1$ approaches a value of about 1.21, the largest eigenvalue becomes positive and the stable solution, as shown in Figure \ref{supfig} (a), gives rise to stable limit cycle encircling a stable fixed point. This indicates to emergence of a supercritical Hopf bifurcation. Figure \ref{supfig} (b) is the bifurcation diagram for $L_r$ for the same range of $c_2$: At $c_1\approx -1.21$, stable fixed point (red), through a supercritical Hopf bifurcation, gives rise to stable limit cycle (green) which in turn generates multiple limit cycles (blue) before reaching a region (cream) where multiple aperiodic trajectories corresponding to different initial conditions reside.
